\documentstyle[amssymb,preprint,eqsecnum,aps]{revtex}

\begin{document}
\draft
\title{Competition of charge, orbital, and ferromagnetic correlations in layered
manganites}
\author{D.B. Romero$^{1,2}$, Y. Moritomo$^{3}$, J.F. Mitchell$^{4}$, and H.D. Drew$%
^{2}$}
\address{$^{1}$NIST - Optical Technology, Gaithersburg, MD 20899\\
$^{2}$Physics Department, University of Maryland, College Park, MD 20742\\
$^{3}$CIRSE and Department of Applied Physics, Nagoya University, Nagoya
464-01, Japan\\
$^{4}$Materials Science Division, Argonne National Laboratory, Argonne, IL
60439}
\date{\today}
\maketitle

\begin{abstract}
The competition of charge, orbital, and ferromagnetic interactions in
layered manganites is investigated by magneto-Raman scattering spectroscopy.
We find that the colossal magnetoresistance effect in the layered compounds
results from the interplay of the orbital and ferromagnetic double-exchange
correlations. Inelastic scattering by charge-order fluctuations dominates
the quasiparticle dynamics in the ferromagnetic-metal state. The scattering
is suppressed at low frequencies, consistent with the opening of a
charge-density wave pseudogap.
\end{abstract}





Quantum criticality, driven by competing electronic ground states, can lead
to dramatic tunability of the physical properties of strongly-correlated
condensed-matter systems \cite{sachdev,orenstein}. A current case of great
interest is the role of collective excitations in the phenomenon of
``colossal'' magnetoresistance (CMR) in manganites. These excitations
represent the various interactions involving the electrons in the Mn {\it d}
orbitals \cite{tokura1}. Charge ordering (CO), where the Mn$^{3+}$/Mn$^{4+}$
ions form an ordered sublattice, occurs because the large on-site Coulomb
repulsion forbids the double-occupancy of the {\it e}$_{g}$ orbitals.
Orbital ordering (OO) of the Mn$^{3+}$ {\it e}$_{g}$ electron orbitals
results from a cooperative Jahn-Teller (JT) distortion induced by
electron-phonon interactions. Antiferromagnetism is associated with the
superexchange coupling of the {\it t}$_{2g}$ electron spins. The
ferromagnetic-metal (FM) ground state is attributed to Mn$^{3+}$---O---Mn$%
^{4+}$ double-exchange, a consequence of strong Hund's coupling in which
parallel spin alignment favors electron hopping between neighboring Mn
atoms. The CMR effect is observed in ferromagnetic metals as a large
decrease in their electrical resistance near the Curie temperature ($T_{C}$)
upon application of a magnetic field ($H$) \cite{rao}. There is a growing
realization that the interplay of these collective excitations is
responsible for the sensitivity of this phenomenon to external perturbations 
\cite{millis2}.

A preponderance of evidence points to the coexistence of the different
electronic phases in the pseudocubic manganites. Thermopower measurements
have shown that the entropy for $T\gg T_{C}$ is smaller than that of an
uncorrelated insulator \cite{mahendiran}. This agrees with a recent x-ray
diffuse scattering study \cite{shimomura} that finds the JT polarons in the
paramagnetic state are correlated. The observation of Mn$^{3+}$/Mn$^{4+}$
striped phases \cite{mori} suggests that the correlation is related to
charge and orbital ordering. Charge-order insulator compounds exhibit a
transition to the FM phase above an $H$ that is small compared to the charge
ordering transition temperature ($T_{co}$) i.e., $g\mu _{B}H\ll k_{B}T_{co}$ 
\cite{tokura2}, indicating a small free energy separates these two ground
states. This could lead to charge ordering persisting with the ferromagnetic
double-exchange correlations below $T_{C}$. Evidence for such coexistence
was reported in earlier infrared studies \cite{kim,okimoto1} which estimate
the itinerant carrier effective mass as $m^{*}/m_{e}\approx 13$ \cite{kim}
or $80$ \cite{okimoto1}. These are anomalously much larger than $%
m^{*}\approx 2.5m_{e}$, the value from specific heat \cite{hamilton}. Within
the charge ordering scenario, the difference can be reconciled by a
charge-density wave opening a partial gap at the Fermi level ({\it E}$_{F}$)
below $T_{C}$, suppressing both the Drude spectral weight and the density of
states at {\it E}$_{F}$. Consequently, the optical mass increases while the
specific heat mass decreases. However, a recent work \cite{simpson} has
shown that the Drude weights were significantly underestimated in the bulk
optical studies \cite{kim,okimoto1}. Measurements on thin films of optimally
doped compounds yielded $m^{*}/m_{e}\approx 3-4$, comparable to the specific
heat mass \cite{simpson}. This result rules out the presence of strong
charge-order correlations in the FM state of the pseudocubic compounds \cite
{simpson}.

In this work, we investigate the competition of the collective ground state
excitations in the layered manganites. The effects of competing ground
states should be more pronounced in these two-dimensional (2D) systems since
the propensity towards a quantum critical point is exacerbated in lower
dimensions \cite{sachdev,orenstein}. Indeed, the CMR effect is enhanced in
the quasi-2D manganites \cite{moritomo1}. We demonstrate that the CMR
phenomenon in these compounds results from the interplay of orbital and
ferromagnetic double-exchange correlations. In contrast with the pseudocubic
materials, we find evidence in the layered manganites for an unusual
quasiparticle dynamics in the FM state arising from strong inelastic
scattering by charge-order fluctuations.

Our investigation was carried out on single-crystals of the double-layer
manganites $R_{2-2x}$Sr$_{1+2x}$Mn$_{2}$O$_{7}$ with the cations $R=$ La or
Nd and for doping $x=0.4$ or $0.5$. These crystals consist of MnO$_{2}$
bilayers separated by insulating ($R$,Sr)$_{2}$O$_{2}$ sheets, a quasi-2D
structure responsible for their anisotropic transport and magnetic
properties \cite{moritomo1,moritomo2}. LaSr$_{2}$Mn$_{2}$O$_{7}$ is an
antiferromagnetic insulator which undergoes charge and orbital ordering
below $T_{co}\simeq 210$ K \cite{li}. Electron diffraction \cite{li}
provides evidence for real-space checkerboard pattern of Mn$^{3+}$ and Mn$%
^{4+}$ ions and coherent JT displacement of the O atoms within the MnO$_{2}$
planes. In NdSr$_{2}$Mn$_{2}$O$_{7}$, this ordered state is suppressed
because the substitution of La by the larger Nd cation results in a chemical
pressure that relaxes the in-plane JT distortions \cite{takata}. La$_{1.2}$Sr%
$_{1.8}$Mn$_{2}$O$_{7}$ exhibits CMR behavior near $T_{C}\simeq 120$ K \cite
{moritomo1} associated with the paramagnetic-insulator to
ferromagnetic-metal transition. Raman spectroscopy was conducted over the
temperatures $T=$ 5 K to 350 K and magnetic-fields $H=$ 0 T to 7.5 T in a
back-scattering geometry using a triple-grating spectrometer with a
liquid-nitrogen cooled CCD detector and the 514.5 nm Ar$^{+}$ laser-line as
excitation. Charge and orbital correlations are probed via the optical
phonons activated by the lowered symmetry due to the ordering within the MnO$%
_{2}$ layers. Polarized Raman scattering was used to identify the phonon
symmetries. For the double-layer manganites, the following scattering
geometries extract the dominant $A_{1g}$ and $B_{1g}$ modes of the $D_{4h}$
point-group: $zz\Rightarrow A_{1g}$, $x^{\prime }x^{\prime }\Rightarrow
A_{1g}+B_{2g}$, $x^{\prime }y^{\prime }\Rightarrow $ $B_{1g}+A_{2g}$ where
the notation {\bf \^{e}}$_{i}${\bf \^{e}}$_{s}$ refers to the polarization
of the incident and scattered light and $z$ is parallel to the $c$-axis
while $x^{\prime }$ and $y^{\prime }$ are along the Mn---O bonds in the MnO$%
_{2}$ plane. Quasiparticle dynamics is inferred from electronic Raman
scattering.

In the uniform state ($T>T_{co}$), shown in Fig. \ref{Figure1}A for $T=296$
K, LaSr$_{2}$Mn$_{2}$O$_{7}$ reveals only the phonons of its tetragonal
crystal structure. This structure has $I4/mmm$ ($D_{4h}^{17}$) symmetry for
which we expect $4A_{1g}+B_{1g}+5E_{g}$ Raman-active normal modes. As
predicted, there are four peaks at 180 cm$^{-1}$, 250 cm$^{-1}$, 460 cm$%
^{-1} $ and 570 cm$^{-1}$ in the $x^{\prime }x^{\prime }$ and $zz$ spectra
corresponding to the $A_{1g}$ modes and one at 325 cm$^{-1}$ in the $%
x^{\prime }y^{\prime }$ for the only allowed $B_{1g}$ mode. To determine the
dominant atomic vibration in these modes, we invoke the four-fold and
two-fold rotational symmetry of the $A_{1g}$ and $B_{1g}$ modes,
respectively. The four-fold axially symmetric {\it c}-axis vibrations of the
(La,Sr), Mn, and apical O$_{c}$ atoms along the tetragonal axis are
respectively assigned to the three $A_{1g}$ phonons at 180 cm$^{-1}$, 250 cm$%
^{-1}$, and 570 cm$^{-1}$. The in-plane O$_{ab}$ atoms have two Raman-active
vibrations along the {\it c}-axis, a four-fold symmetric in-phase motion
attributed to the remaining $A_{1g}$ mode at 460 cm$^{-1}$ and a two-fold
symmetric out-of-phase motion comprising the $B_{1g}$ phonon at 325 cm$^{-1}$%
.

In the charge-ordered state ($T<T_{co}$), new peaks marked by the arrows in
Fig. \ref{Figure1}B for $T=5$ K emerge in the $x^{\prime }x^{\prime }$ and $%
x^{\prime }y^{\prime }$ spectra but not in $zz$. Fig. \ref{Figure1}C
highlights the absence of these features in NdSr$_{2}$Mn$_{2}$O$_{7}$ in
which the ordering is quenched by the La to Nd cation substitution. From
these results, we conclude that the new modes are activated by the lowered
symmetry within the MnO$_{2}$ layers in LaSr$_{2}$Mn$_{2}$O$_{7}$.

In Fig. \ref{Figure1}B, the peak at 250 cm$^{-1}$ in $x^{\prime }y^{\prime }$
(dashed arrow) is a CO activated Mn phonon. Above $T_{co}$, the uniform Mn
charge distribution in the Mn$^{3.5+}$---O$_{ab}$---Mn$^{3.5+}$ bonds
precludes observing this mode. Below $T_{co}$, the ordered Mn sublattice
breaks this symmetry. In this case, the Mn atom vibration modulates the Mn$%
^{3+}$---O$_{ab}$---Mn$^{4+}$ bond polarizability leading to the observed
Raman-activity. The anomalous $B_{1g}$ character of this new Mn mode
reflects the two-fold symmetric checkerboard Mn$^{3+}$/Mn$^{4+}$ pattern in
the CO state.

The remaining new modes (solid arrows) in Fig. \ref{Figure1}B are the
phononic signatures of orbital ordering. The concomitant cooperative JT
distortion of such ordering induces the peaks at 520 cm$^{-1}$ and 636 cm$%
^{-1}$ in $x^{\prime }x^{\prime }$ and at 522 cm$^{-1}$ and 694 cm$^{-1}$ in 
$x^{\prime }y^{\prime }$. Recently, Yamamoto {\it etal.} \cite{yamamoto}
assigned these modes to breathing and JT vibrations of the O$_{ab}$ atoms.
The evidence for $e_{g}$ orbital ordering is presented in Fig. \ref{Figure2}%
. It is suggested by the appearance of weak overtones at twice the frequency
of the JT phonons in the OO ground state of pseudocubic LaMnO$_{3}$,
double-layer LaSr$_{2}$Mn$_{2}$O$_{7}$ and single-layer La$_{1/2}$Sr$_{3/2}$%
MnO$_{4}$. Multiphonon scattering due to anharmonicities is ruled out since
the two-phonon peaks do not correlate with the strongest phonons observed in
these compounds (for instance, the 2$\omega _{o}$ peak is absent in the $zz$
spectrum of LaSr$_{2}$Mn$_{2}$O$_{7}$ in Fig. \ref{Figure1}B and it does not
correspond to the strong peak at 495 cm$^{-1}$ in LaMnO$_{3}$ in Fig. \ref
{Figure2}). Fig. \ref{Figure2} inset depicts a resonance Raman mechanism
adapted from the recent work of Allen and Perebeinos \cite{allen} on LaMnO$%
_{3}$. The lower parabola represents the OO ground state configuration while
the upper parabola is the lowest lying excited state of a self-trap exciton (%
$\Delta _{JT}$) that results from an orbital flip followed by lattice
relaxation via a JT distortion ({\bf Q}$_{JT}$) of the O$_{ab}$ atoms. The
self-localized nature of these electronic states leads to Franck-Condon
vibrational sidebands. An electron excited by an incident photon with $\hbar
\omega =2\Delta _{JT}$ can relax to the various ground state vibrational
levels accounting for the multiphonon features in our Raman spectra. The
relative intensity, $I(2\omega _{o})/I(\omega _{o})$, of the Franck-Condon
peaks is smaller in La$_{1/2}$Sr$_{3/2}$MnO$_{4}$ and LaSr$_{2}$Mn$_{2}$O$%
_{7}$ ( $\simeq 1/8$ ) than in LaMnO$_{3}$ ( $\simeq 1/2$ ). This is
explained by noting that in LaMnO$_{3}$, with only Mn$^{3+}$ ions, the large
on-site Coulomb repulsion suppresses hopping making the electron-phonon
interaction effective in self-trapping the {\it e}$_{g}$ electron \cite
{allen}. This condition is alleviated in the half-filled layered manganites
due to the presence of Mn$^{4+}$ sites resulting in the reduced $2\omega
_{o} $ peak.

In previous works, the magnetic field-induced melting of the charge-order
ground state was inferred only indirectly from the observation of an
antiferromagnetic-insulator to ferromagnetic-metal transition in transport
and magnetization studies \cite{tokura2}. Direct structural evidence for
this effect is given in Fig. \ref{Figure1}B showing the suppression of the
CO and OO features in LaSr$_{2}$Mn$_{2}$O$_{7}$ when $H=7$ T is applied.
Further insight into the origin of the suppression is gained from our
findings in the CMR compound La$_{1.2}$Sr$_{1.8}$Mn$_{2}$O$_{7}$.

The phononic signatures of the charge-ordered state are also observed in La$%
_{1.2}$Sr$_{1.8}$Mn$_{2}$O$_{7}$. In the $xx$ ($\Rightarrow A_{1g}+B_{1g}$)
spectra in Fig. \ref{Figure3}, the CO induced Mn peak is at 240 cm$^{-1}$
while the OO activated O$_{ab}$ phonons are at 514 cm$^{-1}$ and 623 cm$%
^{-1} $. These peaks are progressively quenched either by lowering $T$ below 
$T_{C} $ (Fig. \ref{Figure3}A) or by raising $H$ (Figs. \ref{Figure3} B and
C), proof that the melting of the charge-order ground state is driven by the
ferromagnetic alignment of the Mn spins. The dynamics of the competition of
charge, orbital, and ferromagnetic correlations in La$_{1.2}$Sr$_{1.8}$Mn$%
_{2}$O$_{7}$ is illustrated in Fig. \ref{Figure4}. In Fig. \ref{Figure4}A,
charge ordering sets in near $T\simeq 270$ K and grows on lowering $T$
peaking below $T_{C}$ at $T\simeq 100$ K for $H=0$ T. In contrast, Fig. \ref
{Figure4}B shows that orbital coherence exists in the entire
paramagnetic-insulating state attaining a maximum just above $T_{C}$ at $%
T\simeq 130$ K. With $H=7.5$ T, both charge and orbital correlations drop
precipituously near $T_{C}$. The $T$ and $H$ behaviors of the orbital
correlations are similar to that of the short-range polaron correlations
manifested in the diffuse x-ray and neutron scattering in La$_{1.2}$Sr$%
_{1.8} $Mn$_{2}$O$_{7}$ \cite{doloc}. The correspondence between these
behaviors and the paramagnetic-insulator to ferromagnetic-metal transition
in La$_{1.2} $Sr$_{1.8}$Mn$_{2}$O$_{7}$ is evidence that competing orbital
and ferromagnetic double-exchange interactions around $T_{C}$ is responsible
for the CMR phenomenon in manganites. On the other hand, the presence of the
Mn peak below $T_{C}$ seen in Fig. \ref{Figure4}A indicates that charge
correlations persist against the ferromagnetic instability. An additional
support for this conjecture is our observation in Fig. \ref{Figure3} of an
electronic Raman continuum below $T_{C}$.

Electronic Raman scattering can arise from charge-density fluctuations. This
is expressed in terms of an electronic polarizability, $\chi (\omega ,T)$,
which is obtained from the Raman scattering intensity $I(\omega ,T)$ using
the relation $I(\omega ,T)\propto 
\mathop{\rm Im}%
\chi (\omega ,T)/[1-\exp (-\hbar \omega /k_{B}T)]$. Fig. \ref{Figure3} shows
typical $%
\mathop{\rm Im}%
\chi (\omega ,T)$ derived from our measured spectra. Just below $T_{C}$, $%
\mathop{\rm Im}%
\chi (\omega ,T)$ is seen as a nearly flat background continuum. For $T\ll
T_{C}$ (Fig. \ref{Figure3}A) or upon raising $H$ at $T\lesssim T_{C}$ (Fig. 
\ref{Figure3}B and C), $%
\mathop{\rm Im}%
\chi (\omega ,T)$ is suppressed below $\omega \simeq 300$ cm$^{-1}$. While
the pseudocubic manganites manifest a similar featureless $%
\mathop{\rm Im}%
\chi (\omega ,T)$ in the FM state, the low-frequency suppression is not
observed \cite{yoon}. This difference is attributed to an effect of the
lower dimensionality in the layered compounds. It signals a strong inelastic
scattering by persistent CO fluctuations that develop into a collective
charge-density wave excitation below $T_{C}$. This picture is also suggested
by other studies. Optical conductivity \cite{ishikawa} betrays a small Drude
contribution implying an optical mass that is much larger than that derived
by a specific heat study \cite{okuda}. These are indicative of a
charge-density wave pseudogap opening at {\it E}$_{F}$, in accord with the
severe depression of the quasiparticle states at {\it E}$_{F}$ revealed by
an angle-resolved photoemission study \cite{dessau} and suggested by the
re-entrant insulating state below $T\simeq 50$ K invariably seen in
transport measurements \cite{moritomo1,okuda}. It is interesting to note
that the latter temperature is comparable to the onset frequency where $%
\mathop{\rm Im}%
\chi (\omega ,T)$ is suppressed.

In conclusion, competing collective groundstate excitations is a ubiquituous
phenomenon in the manganites whose understanding is crucial in unraveling
the wide tunability of their physical properties. Our observation of an
unusual quasiparticle dynamics below $T_{C}$ in La$_{1.2}$Sr$_{1.8}$Mn$_{2}$O%
$_{7}$ is reminiscent of the unconventional behavior of the normal state of
high-temperature superconducting cuprates \cite{orenstein}, suggesting that
the FM ground state in the layered manganites is another example of a
non-Fermi liquid.

We acknowledge A.J. Millis for elucidating the importance of charge ordering
in the CMR manganites and P.B. Allen for pointing out the Franck-Condon
model of the orbital excitations. We thank Dr. R.V. Datla of NIST-Optical
Technology for his support of this work. JFM acknowledges the U.S. DOE,
Office of Basic Energy Sciences - Materials Sciences Division, under
Contract No. W-31-109-ENG-38.

\begin{figure}[tbp]
\caption{A. For $T>T_{co}$, only the phonons of the tetragonal crystal
structure of LaSr$_{2}$Mn$_{2}$O$_{7}$ are observed. B. For $T<T_{co}$, new
modes in $x^{\prime }x^{\prime }$ and $x^{\prime }y^{\prime }$ are activated
by the charge (dashed arrow) and orbital (solid arrow) ordering within the
MnO$_2$ plane. Magnetic field quenches these modes. C. The new modes are
absent in NdSr$_{2}$Mn$_{2}$O$_{7}$.}
\label{Figure1}
\end{figure}

\begin{figure}[tbp]
\caption{Franck-Condon effect in manganites with orbital-ordered ground
state.}
\label{Figure2}
\end{figure}

\begin{figure}[tbp]
\caption{Persistence of charge (dashed arrow) and orbital (solid arrow)
correlations in La$_{1.2}$Sr$_{1.8}$Mn$_{2}$O$_{7}$ and their melting by
lowering the temperature (A) or raising the magnetic field (B and C). An
electronic background continuum below $T_C$ is suppressed at low frequencies.
}
\label{Figure3}
\end{figure}

\begin{figure}[tbp]
\caption{Temperature and magnetic field dependence of the normalized
intensity of the (A) CO activated Mn and (B) OO induced O$_{ab}$ phonons in
La$_{1.2}$Sr$_{1.8}$Mn$_{2}$O$_{7}$.}
\label{Figure4}
\end{figure}

\end{document}